\documentclass[aps, prd, twocolumn, preprintnumbers,nofootnoteinbib]{revtex4-1}
\usepackage[T1]{fontenc}
\usepackage{microtype}
\usepackage[textsize=scriptsize,backgroundcolor=red!70,linecolor=red]{todonotes}
\usepackage{booktabs}
\usepackage{dcolumn}
\usepackage{amsmath}
\usepackage{amsfonts}
\usepackage{amssymb}
\usepackage{graphicx}
\usepackage{hyperref}
\usepackage[all]{hypcap}
\usepackage{paralist}
\usepackage{multirow}

\newcommand{\chisq}{\ensuremath{\chi^2}}

\usepackage{graphicx}
\usepackage{dcolumn}
\usepackage{bm}
\usepackage{epsf}
\usepackage{dcolumn}
\def\be{\begin{equation}}
\def\ee{\end{equation}}
\def\bea{\begin{eqnarray}}
\def\eea{\end{eqnarray}}
\def\gsim{\ \rlap{\raise 2pt\hbox{$>$}}{\lower 2pt \hbox{$\sim$}}\ }
\def\lsim{\ \rlap{\raise 2pt\hbox{$<$}}{\lower 2pt \hbox{$\sim$}}\ }
\def\dslash{\kern-4pt \not{\hbox{\kern-2pt $\partial$}}}
\def\pslash{\not{\hbox{\kern-2pt p}}}

\begin{document}
\DeclareGraphicsExtensions{.eps,.ps}


\title{Impact of nonstandard interactions on sterile neutrino searches at IceCube}



\author{Jiajun Liao}
\affiliation{Department of Physics and Astronomy, University of Hawaii at Manoa, Honolulu, HI 96822, USA}
 
\author{Danny Marfatia}
\affiliation{Department of Physics and Astronomy, University of Hawaii at Manoa, Honolulu, HI 96822, USA}

\begin{abstract}

We analyze the energy and zenith angle distributions of the latest 2-year IceCube dataset of upward going atmospheric neutrinos to constrain sterile neutrinos at the eV scale in the $3+1$ scenario. We find that the parameters favored by a combination of LSND and MiniBooNE data are excluded at more than the
99\%~C.L. We explore the impact of nonstandard matter interactions on this exclusion, and find that the exclusion holds for nonstandard interactions (NSI) that are within the stringent
model-dependent bounds set by collider and neutrino scattering experiments. However, for large NSI parameters subject only to model-independent bounds from neutrino oscillation experiments, the LSND and MiniBooNE data are consistent with IceCube.
  
\end{abstract}
\maketitle

The discovery of neutrino oscillations is a significant triumph. The results of most neutrino experiments  can be successfully explained in the framework of the
standard model (SM) with three massive neutrinos~\cite{Agashe:2014kda}. However, anomalies in short baseline experiments hint at physics beyond the three-neutrino framework. The first anomaly emerged from the Liquid Scintillator Neutrino Detector (LSND) experiment, which found $3.3 \sigma$ C.L. evidence for $\bar{\nu}_\mu\rightarrow\bar{\nu}_e$ oscillations with a mass-squared difference $\Delta m^2\sim 1 \text{ eV}^2$~\cite{Aguilar:2001ty}. A search by the Mini-Booster Neutrino Experiment (MiniBooNE) with similar $L/E\sim 1$ m/MeV but in a different energy range and distance found an excess in the low-energy regions of both the electron and anti-electron neutrino events~\cite{Aguilar-Arevalo:2013pmq}, which is consistent with the results from LSND. The two experiments together suggest that there exist sterile neutrinos at a mass scale of 1~eV that mix with the standard three neutrinos.

As atmospheric muon neutrinos propagate through the earth, oscillations with eV-mass sterile neutrinos undergo resonance enhancement at a TeV~\cite{Nunokawa:2003ep} due to matter effects~\cite{Wolfenstein:1977ue,Mikheev:1986gs}. This leads to a distortion of the energy and zenith angle distributions of the muon track events at IceCube thereby providing a crucial test of the LSND/MiniBooNE anomaly. Studies of sterile neutrinos in early IceCube configurations can be found in Refs.~\cite{Razzaque:2011ab, Esmaili:2013vza, Lindner:2015iaa}. The latest 
2-year dataset from the 79-string and 86-string IceCube configurations is comprised of 35,000 upward going muon neutrino events and can be found in Refs.~\cite{Aartsen:2015rwa, IceCubedata}. One year of IceCube-86 data have been utilized to search for sterile neutrinos in Ref.~\cite{Jones:2015bya}.

In this Letter, we study the effects of nonstandard interactions (NSI) in neutrino propagation on sterile neutrino searches at IceCube. We consider the simplest $3+1$ mass scheme, with an eV-mass sterile neutrino.

NSI are motived by physics beyond the SM, and their effects on neutrino oscillations have been extensively studied; for a review see Ref.~\cite{Ohlsson:2012kf}. Similar to the standard matter effect, NSI in matter have a large effect on atmospheric neutrinos traveling through the earth due to coherent interactions. Matter NSI can be described in an effective theory by the dimension-six operators~\cite{Wolfenstein:1977ue}
\be
  \label{eq:NSI}
  \mathcal{L}_\text{NSI} =2\sqrt{2}G_F
   \epsilon^{fC}_{\alpha\beta} \!
        \left[ \overline{\nu_\alpha} \gamma^{\rho} P_L \nu_\beta \right] \!\!
        \left[ \bar{f} \gamma_{\rho} P_C f \right] + \text{h.c.}\,,
\ee
where $\alpha, \beta=e, \mu, \tau$, $C=L,R$, $f=u,d,e$, and
$\epsilon^{fC}_{\alpha\beta}$ are dimensionless parameters that define the
strength of the new interaction in units of the Fermi constant $G_F$.

Matter NSI parameters can be constrained by collider and neutrino scattering experiments.
 The ${\cal O}(0.01)$ to ${\cal O}(0.1)$ bounds obtained are model dependent because they typically assume mediator masses heavier than ${\cal O}(100)$~GeV~\cite{Biggio:2009kv}. However, for 10~MeV-mass mediators, the bounds can be relaxed to ${\cal O}(1)$~\cite{Farzan:2015hkd}. Model independent bounds on matter NSI parameters mainly arise from neutrino oscillation data (since integrating out the mediator in the $t$-channel forward scattering amplitude leads to a contact interaction irrespective of the mediator mass) and are derived in the three-neutrino framework; see e.g., Ref.~\cite{Gonzalez-Garcia:2013usa}. Because three-neutrino oscillations are not sensitive to an overall diagonal NSI parameter, bounds are often set on the differences of diagonal NSI parameters. It is therefore possible to have very large values for the diagonal NSI parameters with small differences between them. In what follows, we only consider nonstandard interactions of active neutrinos, and assume that the sterile neutrino has no nonstandard interactions. 

{\bf Survival probabilities.}
The unitary matrix that mixes the mass eigenstates $\nu_i$ ($i=1,2,3,4$) with the flavor eigenstates $\nu_\alpha$ ($\alpha=e,\mu,\tau,s$) is
$U=R_{34}V_{24}V_{14}R_{23}V_{13}R_{12}$,
where $R_{ij}$ is a real rotation by an angle $\theta_{ij}$ in the $ij$ plane, and $V_{ij}$ is a complex rotation by $\theta_{ij}$ and a phase $\delta_{ij}$.  For IceCube neutrinos with energy above 500~GeV, the $\nu_e$ flavor can be neglected because the atmospheric $\nu_e$ flux is small compared to the $\nu_\mu$ flux and because $\nu_e$ mixing is suppressed except in the resonance region. Also, the mass splittings between active neutrinos are negligible. We set all the phases in the mixing matrix to zero, and assume the NSI parameters to be real. With these simplifications, the Hamiltonian that describes
the propagation of the three-flavor system of atmospheric neutrinos in matter is
\bea
H&=&\frac{\Delta m_{41}^2}{2E_\nu}\left[\begin{pmatrix}
   0 & s_{24}s_{34} & s_{24}c_{34} \\
   s_{24}s_{34} & s^2_{34} & s_{34}c_{34} \\
   s_{24}c_{34} & s_{34}c_{34} & c^2_{34}
   \end{pmatrix}\right.
\nonumber\\
 &+&\left. \hat{A} \begin{pmatrix}
    \epsilon_{\mu\mu} & \epsilon_{\mu\tau} & 0 \\
    \epsilon_{\mu\tau} & \epsilon_{\tau\tau} & 0 \\
    0 & 0 & \kappa
    \end{pmatrix} \right] + {\cal O}(s_{14}^2 , s_{24}^2)\,,
\label{eq:H}
\eea
where $\Delta m_{41}^2=m_4^2-m_1^2$, $c_{ij}$ ($s_{ij}$) denotes $\cos\theta_{ij}$ ($\sin\theta_{ij}$), $\hat{A}=\frac{2E_\nu V_{CC}}{\Delta m_{41}^2}$, $V_{CC}=\sqrt{2}G_FN_e$ is the electron charged-current potential, $\kappa=\frac{N_n}{2N_e}\simeq 0.5$ is the ratio of the standard neutral-current interaction to the charged-current interaction, $\epsilon_{\alpha\beta}\equiv\sum\limits_{f,C}\epsilon^{fC}_{\alpha\beta}\frac{N_f}{N_e}$ parametrize the strength of NSI relative to the SM charged-current interaction in matter, and $N_f$ is the number density of fermion $f$. 

\begin{figure}
\includegraphics[width=0.23\textwidth]{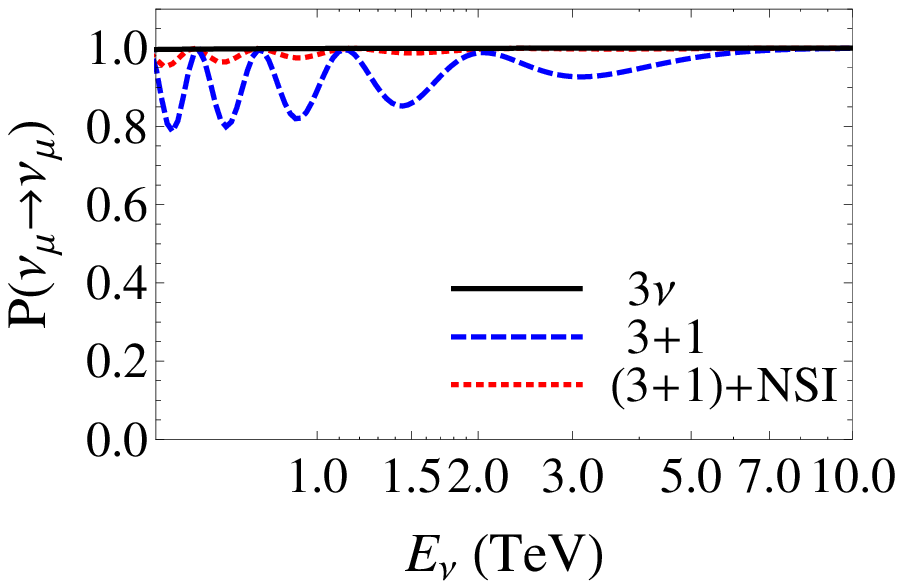}
\includegraphics[width=0.23\textwidth]{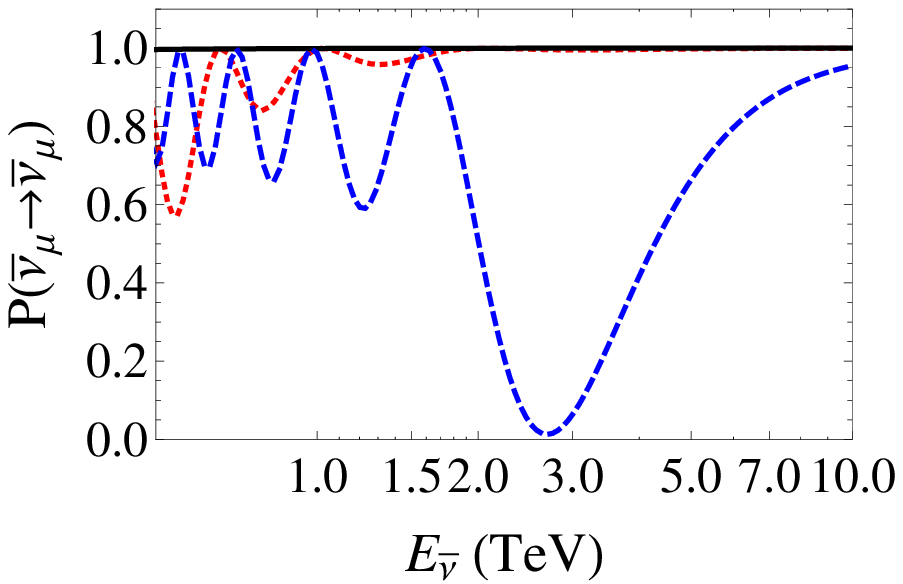}
\caption{The survival probability for $\nu_\mu$  and $\bar{\nu}_\mu$  for atmospheric neutrinos travelling through the earth with zenith angle $\cos \theta_z=-0.8$. The black solid lines show the standard $3\nu$ oscillations with the best-fit values from Ref.~\cite{Agashe:2014kda}. The blue dashed curves correspond to the $3+1$ sterile neutrino scenario with $\sin^2 \theta_{14}=0.023$, $\sin^2 2\theta_{24}=0.25$ and $\Delta m_{41}^2=0.63 \text{ eV}^2$. The red dotted curves correspond to the $3+1$ scenario with NSI parameters $\epsilon_{\mu\mu}=-6.26$ and $\epsilon_{\tau\tau}=-6.4$. All other parameters are set to zero.}
\label{fig:prob}
\end{figure}

The survival probabilities can be calculated numerically using the GLoBES 
software~\cite{GLOBES} supplemented with the new physics tools of Ref.~\cite{Kopp:2006wp}.
An illustration of the survival probabilities that uses the density profile of the Preliminary Reference Earth Model~\cite{Dziewonski:1981xy} and shows the resonance in the antineutrino channel (since $\Delta m^2_{41}>0$) can be found in Fig.~\ref{fig:prob}.
In order to understand the dependence of the survival probabilities on the NSI parameters, we assume a constant matter density for simplicity, and define $M=\frac{2E_\nu H}{\Delta m_{41}^2}-\epsilon_{\mu\mu}\hat{A}I_3$, which is diagonalized by a mixing matrix $U'$. The $\nu_\mu$ survival probability is
\be
P_{\nu_\mu\nu_\mu}=1-4\sum_{j<k}|U'_{\mu j}|^2 |U'_{\mu k}|^2 \sin^2 (\lambda_k-\lambda_j)\frac{\Delta m_{41}^2 L}{4E_\nu}\,,
\label{eq:prob}
\ee
where $\lambda_j$ ($j=1,2,3$) are the eigenvalues of $M$. For $|\epsilon_{\mu\tau}|,
|\epsilon_{\mu\mu}-\epsilon_{\tau\tau}|,s_{24} \ll 1$, we use perturbation theory~\cite{pert} to find
\bea
U'_{\mu 1}&\simeq& 1\,,
\nonumber \\
U'_{\mu 2}&\simeq& \frac{2[s_{24}\sin(\theta_{34}-\xi)+\epsilon_{\mu\tau}\hat{A}\cos\xi]}{\lambda_2-\lambda_1}\,,
\nonumber \\
U'_{\mu 3}&\simeq& \frac{2[s_{24}\cos(\theta_{34}-\xi)+\epsilon_{\mu\tau}\hat{A}\sin\xi]}{\lambda_3-\lambda_1}\,,
\label{eq:U}
\eea
where $\xi=\frac{1}{2}\arctan\frac{\sin 2\theta_{34}}{\cos 2\theta_{34}+(\kappa-\epsilon_{\tau\tau})\hat{A}}$, and $\lambda_1\simeq 0$,
\bea
\lambda_{2,3}&\simeq&\frac{1}{2}\left[1+(\kappa-\epsilon_{\tau\tau})\hat{A}\right.
\nonumber\\
&\mp&\left.\sqrt{1+2\cos2\theta_{34}(\kappa-\epsilon_{\tau\tau})\hat{A}+(\kappa-\epsilon_{\tau\tau})^2\hat{A}^2}\right].
\label{eq:lambda}
\eea
For antineutrinos, $\hat{A}\rightarrow-\hat{A}$. As we show below, IceCube data are consistent with standard $3\nu$ oscillations, for which the survival probability is very close to unity for $E_\nu \gtrsim 500$~GeV.  
Since the data cover a wide range of energies and oscillation lengths, deviations of the survival probabilities from unity are mainly governed by the values of $|U'_{\mu j}|$. Hence, from Eq.~(\ref{eq:U}) we expect nonzero values of $\epsilon_{\mu\tau}$ to be strongly constrained. Also, since $\lambda_{2,3}$ depend on $\epsilon_{\tau\tau}\hat{A}$ and $\hat{A}$ is proportional to $E_\nu$, $|U'_{\mu j}|$ could get resonantly enhanced as $\lambda_2$ or $\lambda_3$ approaches zero for a particular $E_\nu$. However, for large $\epsilon_{\tau\tau}$, $|U'_{\mu j}|$ will be suppressed and the survival probabilities will be close to unity. Thus we expect that it will be difficult for IceCube data to exclude sterile neutrino scenarios with large values of $\epsilon_{\tau\tau}$.

{\bf IceCube detector simulation.}
The observables of interest at IceCube are the energy and direction of the (anti)muons from $\nu_\mu$ ($\bar{\nu}_\mu$) charged-current interactions. For muon track events, the angular resolution is reported to be less than $1^\circ$~\cite{Weaver}, and since the angle between the neutrino and muon momenta is negligible for high energy events, we safely ignore the difference between the zenith angle of the neutrino and the reconstructed zenith angle of the muon. However, the IceCube detector has poor energy resolution. 
Since the majority of muon events with TeV energies are not fully contained within the instrumented volume of the detector, the energy measured could be arbitrarily smaller than the initial muon energy. The average photon density along the muon track (i.e., the energy loss observed in the detector) is used as a proxy for the muon energy.
The {\it muon energy proxy} is computed by fitting the amount of light expected from the emission of a template muon to the number of observed photons in each event~\cite{Aartsen:2013vja}. Although the energy proxy is only loosely connected to the true neutrino energy, it is a useful
statistical tool. 

The expected number of observed muon track events with the reconstructed muon energy proxy in the range $[ E_{\mu}^{proxy},   E_{\mu}^{proxy}+\Delta_j ( E_{\mu}^{proxy})]$ and zenith angle in the range $[\cos\theta_z, \cos\theta_z+\Delta_i (\cos\theta_z)]$ is given by~\cite{IceCubedata}
\bea
&&N_{ij}^{th} =
\sum_{y} T_y \int_{\Delta_i (\cos\theta_z)} d\cos\theta_z \int_{\Delta_j ( E_{\mu}^{proxy})} d E_{\mu}^{proxy} \int dE_\nu  
\nonumber\\
&&  \eta(E_{\mu}^{proxy}, E_\nu, \cos\theta_z; y) A_{eff} (E_{\mu}^{proxy}, E_\nu, \cos\theta_z; y)
\nonumber\\ 
&&\qquad \times \left[P_{\nu_\mu\nu_\mu}(E_\nu, \cos\theta_z)\Phi_{\nu_\mu}(E_\nu, \cos\theta_z)\right]+(\nu\rightarrow\bar{\nu}) \,,
\label{eq:Nthij}
\eea
where $\Phi_{\nu_\mu}(E_\nu, \cos\theta_z)$ is the atmospheric $\nu_\mu$ flux at the surface of the earth~\cite{Honda:2006qj} as modified by the IceCube collaboration~\cite{Aartsen:2015rwa}, $P_{\nu_\mu\nu_\mu}(E_\nu, \cos\theta_z)$ is the muon neutrino survival probability at the IceCube detector, $A_{eff}$ is the neutrino effective area, and $\eta$ is the optical efficiency of the detector modules, which depends on the true neutrino energy spectrum~\cite{Weaver}. The values of $A_{eff}$ and $\eta$ for the 79-string and 86-string detector configurations are available in Ref.~\cite{IceCubedata}. The livetimes $T_y$ of the two data recording periods are 315.8 and 343.7 days, respectively.

\begin{figure}
\includegraphics[width=0.45\textwidth]{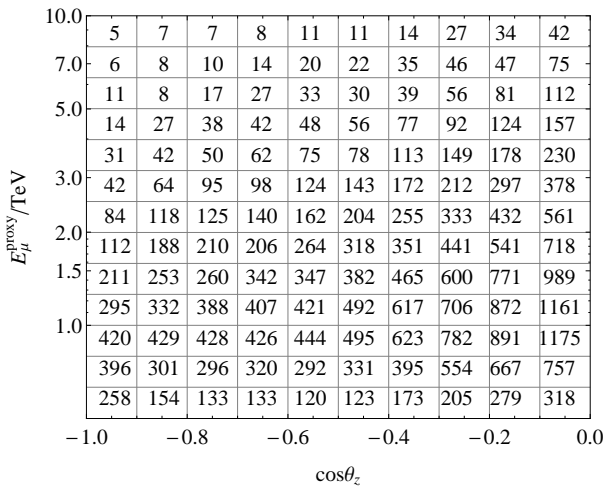}
\caption{Event distribution as a function of the muon energy proxy and $\cos \theta_z$. The event counts per bin are extracted from the 2 year Icecube dataset in Ref.~\cite{IceCubedata}.}
\label{fig:box}
\end{figure}

\begin{figure}
\includegraphics[width=0.23\textwidth]{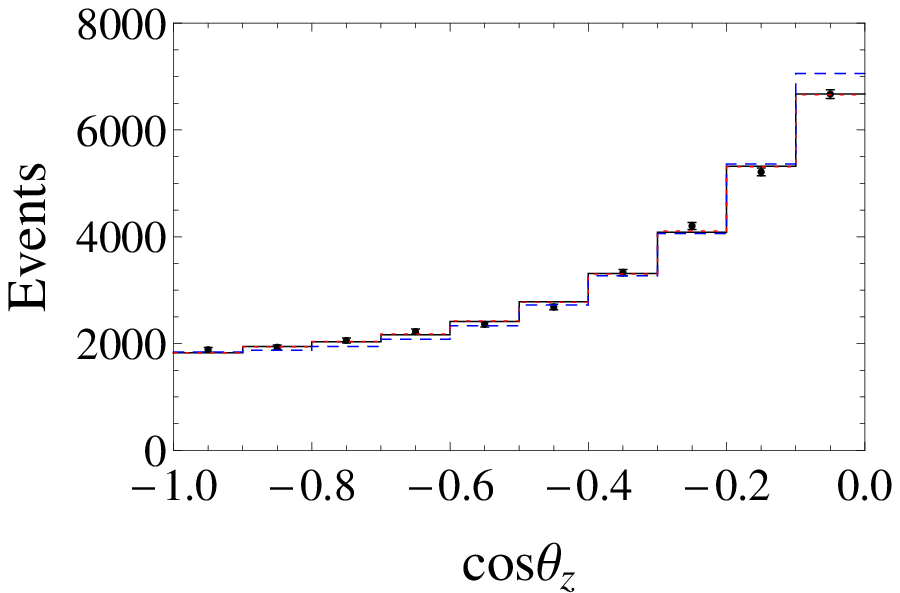}
\includegraphics[width=0.23\textwidth]{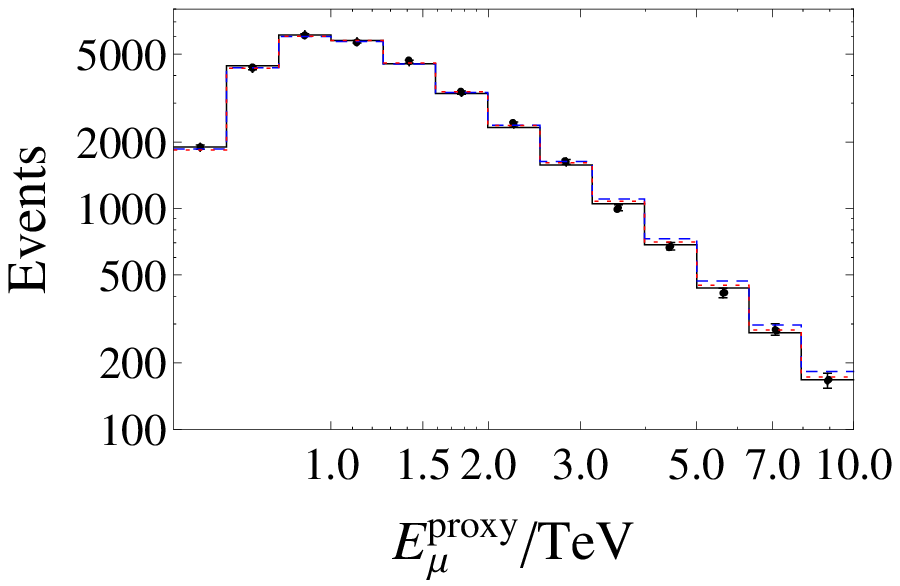}
\caption{Zenith angle and energy distributions of upward going muon events in the 2-year IceCube dataset. The theoretical distributions are for the $3\nu$, $3+1$ and $(3+1)+$NSI scenarios with the parameters and line types as in Fig.~\ref{fig:prob}, and flux normalization factors 1.009, 1.118 and 1.038, respectively.}
\label{fig:distribution}
\end{figure}

{\bf Analysis.}  We choose the same binning edges in our analysis as that in Ref.~\cite{IceCubedata}, and consider 13 bins in the energy range 501~GeV~$\le E_{\mu}^{proxy} \le 10$~TeV and 10~bins in the zenith angle range $-1 \le \cos\theta_z \le 0$. The observed event counts per bin extracted from the 2-year IceCube data are shown in Fig.~\ref{fig:box}. The $\cos \theta_z$ and $E_{\mu}^{proxy}$ event distributions are shown in Fig.~\ref{fig:distribution}. One can see from Fig.~\ref{fig:distribution} that suitably chosen NSI parameters can reconcile discrepant $3+1$ oscillations with IceCube data.

Since the muon neutrino survival probabilities are very close to unity for standard 3$\nu$ oscillations, to calculate the expected number of events for nonstandard oscillations, we simply modify the conventional atmospheric flux by multiplying it with the corresponding survival probability. Note that although we choose $E_{\mu}^{proxy} >501$~GeV, it is possible that $E_\nu $ is less than 500~GeV due to experimental misidentification (see e.g., Suppl. Fig. 4(b) in Ref.~\cite{Aartsen:2015rwa}). In this case, the survival probability for 3$\nu$ oscillations is a little smaller than unity, but since the probability of having $E_\nu <500$~GeV for $E_{\mu}^{proxy} >501$~GeV is very small, it has little effect on our results. 

To evaluate the statistical significance of a nonstandard oscillation scenario, we define 
\be
\chisq =\frac{(1-\alpha)^2}{\sigma_\alpha^2} + \sum_{i=1}^{10} \sum_{j=1}^{13} 2(\alpha N_{ij}^{th} - N_{ij}^{obs} + N_{ij}^{obs}\ln\frac{N_{ij}^{obs}}{\alpha N_{ij}^{th}})\,,\nonumber
\ee
where $\sigma_\alpha=25\%$ is the percent uncertainty in the atmospheric flux normalization~\cite{Honda:2006qj}, $N_{ij}^{obs}$ is the observed event counts per bin from Fig.~\ref{fig:box}, and $N_{ij}^{th}$ is the expected number of events per bin calculated using Eq.~(\ref{eq:Nthij}). We choose the standard $3\nu$ oscillation parameters to be the best-fit values in Ref.~\cite{Agashe:2014kda}, so that the $\chisq$ for the $3\nu$ oscillation scenario is only a function of the normalization factor $\alpha$. We find $\chisq_\text{min,$3\nu$}=112$ with $\alpha=1.009$.

For the $3+1$ scenario, we fix $\sin^2\theta_{14}=0.023$, which is the best fit value from an analysis of reactor neutrino disappearance data~\cite{Kopp:2013vaa}. Although IceCube data are not sensitive to  $\theta_{14}$, we select this nonzero value to be consistent with LSND and MiniBooNE data which measure a nonzero value for the oscillation amplitude $\sin^22\theta_{14}\sin^2\theta_{24}$; the relatively large best fit value for $\theta_{14}$ permits correspondingly small values of $\theta_{24}$ to explain the LSND/MiniBooNE data, thus making the best fit $\theta_{14}$ a conservative choice. Also, we conservatively fix $\theta_{34}=0$, which weakens the sterile neutrino signal at IceCube~\cite{Esmaili:2013vza}. It is noteworthy that for  $\theta_{34}=90^\circ$, sterile neutrino oscillations through the earth occur as though in vacuum which yields an even weaker signal than for \mbox{$\theta_{34}=0$} in some regions of parameter space~\cite{Lindner:2015iaa}. However, from Eq.~(\ref{eq:H}), we see that vacuum oscillations also result for $\theta_{34}=0$ (and any other value of $\theta_{34}$) if $\epsilon_{\mu\tau}=0$ and $\epsilon_{\mu\mu}=\epsilon_{\tau\tau}=\kappa$. Since we scan over $\epsilon_{\mu\mu}$ and $\epsilon_{\tau\tau}$, we focus on $\theta_{34}=0$. 
We find that the fit for nonzero $\epsilon_{\mu\tau}$ is usually worse than for $\epsilon_{\mu\tau}=0$ (as expected from our discussion after Eq.~\ref{eq:lambda}). In the cases that nonzero $\epsilon_{\mu\tau}$ gives a better fit than $\epsilon_{\mu\tau}=0$, the $\chi^2$ is very marginally smaller with $\epsilon_{\mu\tau}$ very close to 0.
We therefore set $\epsilon_{\mu\tau}=0$.
So our $\chisq$ function for the $3+1$ scenario with NSI depends on $\sin^2 2\theta_{24}$, $\Delta m^2_{41}$, $\epsilon_{\mu\mu}$, $\epsilon_{\tau\tau}$ and  $\alpha$. 
We marginalize over  $\epsilon_{\mu\mu}$, $\epsilon_{\tau\tau}$ and $\alpha$ for each point in the $(\sin^2 2\theta_{24},\Delta m^2_{41})$ plane, and calculate $\Delta \chisq = \chisq_\text{min}-\chisq_\text{min,$3 \nu$}$ for the $3+1$ scenarios. 

\begin{figure*}[t]
	\centering
	\includegraphics[width=1\textwidth]{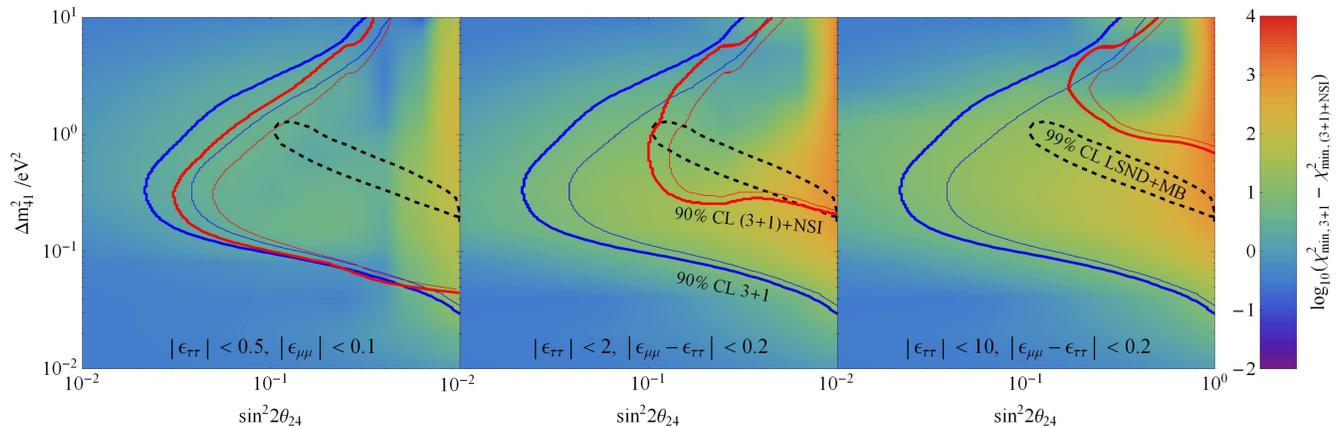}
	\caption{The 90\% and 99\% C.L. exclusion bounds for the $3+1$ scenario from IceCube data are shown in blue; the regions to the right of the curves are excluded. The corresponding bounds for the $3+1$ scenario with NSI parameters in three different ranges are shown in red; all other NSI parameters are set to zero. The black dashed curve encloses the 99\% C.L. allowed region for the combined LSND and MiniBooNE appearance analysis with $\sin^2\theta_{14}=0.023$~\cite{Kopp:2013vaa}. The shading shows the effect of NSI on $3+1$ oscillations. }
	\label{fig:chi}
\end{figure*}

{\bf Results.} In Fig.~\ref{fig:chi}, we display exclusion bounds in the $(\sin^2 2\theta_{24},\Delta m^2_{41})$ plane. To visualize the effect of NSI for a particular pair of sterile neutrino parameters, we also show the values of $\log_{10}( \chisq_\text{min,3+1}-\chisq_\text{min,(3+1)+NSI})$ for each point in the plane. The shading confirms that NSI effects are large for $0.1 \text{ eV}^2<\Delta m^2_{41}< 1 \text{ eV}^2$, and increase as $\sin^2 2\theta_{24}$ is increased.
We see that the LSND and MiniBooNE allowed region is excluded at more than the 99\% C.L. if there is no NSI. 
The impact of NSI on the exclusion bounds depends on the ranges allowed for $\epsilon_{\mu\mu}$ and $\epsilon_{\tau\tau}$. For NSI parameters allowed by collider and neutrino scattering experiments using charged leptons (roughly $|\epsilon_{\tau\tau}|<0.5,
|\epsilon_{\mu\mu}|<0.1$)~\cite{Ohlsson:2012kf}, the exclusion bounds with NSI are only slightly weaker than those without NSI, and the LSND and MiniBooNE allowed region remains excluded at the 99\% C.L.; see the left panel of Fig.~\ref{fig:chi}. This is because NSI effects have an energy dependence different from that of sterile neutrino oscillations. A cancellation of these effects may occur at one energy, whereas the data span a wide energy range. 

Since the bounds from collider and neutrino scattering experiments are model dependent, we also consider model-independent bounds from neutrino oscillation experiments. 
We keep $|\epsilon_{\mu\mu}-\epsilon_{\tau\tau}|<0.2$~\cite{Gonzalez-Garcia:2013usa},
but allow $\epsilon_{\tau\tau}$ to vary in a wide range. We find that as the diagonal NSI parameters become larger, the exclusion bounds on $\sin^2 2\theta_{24}$ for large $\Delta m^2_{41}$ get weaker; see Fig.~\ref{fig:chi}. This can be understood as follows. In the survival probabilities, the NSI parameters are multiplied by $\hat{A}$, which is inversely proportional to $\Delta m^2_{41}$. Hence a large $\Delta m^2_{41}$ can be compensated by a large NSI parameter, thereby suppressing the deviation of the survival probability from unity. In particular, we find that for $|\epsilon_{\tau\tau}| \sim 10$, the combined LSND and MiniBooNE allowed region is no longer excluded by the IceCube data; see the right panel of Fig.~\ref{fig:chi}. Note that such  large values of $|\epsilon_{\tau\tau}|$ do not affect the LSND and MiniBooNE experiments because of their short baselines and low energies. 

{\bf Summary.} We have shown that atmospheric neutrino data from IceCube exclude the simplest $3+1$ sterile neutrino model that explains the LSND and MiniBooNE anomalies at more than the 99\% C.L. However, if nonstandard matter interactions of active neutrinos are sufficiently large ($|\epsilon_{\mu\mu}|\simeq|\epsilon_{\tau\tau}|\simeq 10$), the IceCube bound is completely evaded.  Since in earth matter, $N_u/N_e\simeq N_d/N_e\simeq 3$, NSI parameters $|\epsilon_{\alpha\alpha}| \simeq 10$  could correspond to $|\epsilon_{\alpha\alpha}^u|= 
|\epsilon_{\alpha\alpha}^d|\simeq 1.7$ and $|\epsilon_{\alpha\alpha}^e| \simeq 0$ if we require $\epsilon_{ee}=\epsilon_{\mu\mu}=\epsilon_{\tau\tau}$ (where $\epsilon_{\alpha\alpha}^f=\epsilon_{\alpha\alpha}^{fL}+\epsilon_{\alpha\alpha}^{fR}$, $\alpha=e,\mu,\tau$ and $f=u,d,e$), values that are compatible with constraints from the muon magnetic dipole moment, supernova cooling, meson decays, and fixed target experiments~\cite{Farzan:2015hkd}. (Note that if the $3\times3$ submatrix of NSI parameters in the Hamiltonian is proportional to the identity, the nonstandard interaction can be attributed entirely to the sterile neutrino by subtracting an overall multiple of the identity.) Such large NSI parameters will not be probed at neutrino oscillation experiments like DUNE so long as $|\epsilon_{\mu\mu}-\epsilon_{\tau\tau}| \lsim 0.2$~\cite{Coloma:2015kiu}. A proposed muon-nucleon scattering experiment at the CERN SPS  may provide evidence for dark vector bosons that mediate the NSI (using methods akin to searches for invisible decays of dark photons and mesons)~\cite{sps}.


{\it Acknowledgments.} We thank B.~Jones, J.~Salvado and C.~Weaver for helpful correspondence. This research was supported by the
U.S. DOE under Grant No. DE-SC0010504.

\vspace{0.1 in}
{\bf Note added.} After the completion of this work, the IceCube Collaboration reported the results of their search for sterile neutrinos using one year of IceCube-86 data~\cite{TheIceCube}. Accounting for the fact that we analyzed two years of IceCube data, their exclusion bounds for $3+1$ oscillations without NSI agree very well with the corresponding bounds in our Fig.~\ref{fig:chi}.

\vskip1cm


\end{document}